# Launch Power Optimization for Dynamic Elastic Optical Networks over C+L Bands

F. Arpanaei[(1),(4)], M. Ranjbar Zefreh[(2)], J. A. Hernández[(1)], B. Shariati[(3)], J. Fischer[(3)], J. M. Rivas-Moscoso[(4)], F. Jiménez[(4)], J. P. Fernández-Palacios[(4)], D. Larrabeiti[(1)]

(1) Department of Telematic Engineering, Universidad Carlos III de Madrid, 28911, Leganes, Madrid, Spain,
(2) CISCO Systems S.R.L., Vimercate (MB), Italy,
(3) Fraunhofer Institute for Telecommunications Heinrich Hertz Institute, Einsteinufer 37, 10587 Berlin, Germany,
(4) Telefónica Research and Development, Madrid, Spain, farhad.arpanaei@uc3m.es

*Abstract*—We propose an algorithm for calculating the optimum launch power over the entire C+L bands by maximizing the cumulative link GSNR of a channel plan built upon multiple modulation formats, with application to dynamic EONs. Exact last-fit spectrum assignment proves to outperform exact first-fit in terms of average GSNR at arrival time.

## I. Introduction

The migration of optical networks from C to C+L bands is not easily realized and requires overcoming many operational challenges at different network planes [1]. These challenges are aggravated in elastic optical networks (EONs), where different multi-level modulation format cardinalities (MFC) are used to maximize the spectral efficiency according to the lightpath OSNR penalty. In particular, channel power allocation becomes a burdensome task in operating meshed C+L EONs with dynamic traffic demands, due to stimulated Raman scattering (SRS). Earlier studies addressed the launch power optimization problem in static networks. In [2], it was shown, for the case of uniform launch power, how the channel power could affect the fill margin in C+L static EONs with multiple MFCs, but the overall optimum launch power and reach distance for the entire C+L-band channel plan were not determined. In [3-9], different studies on channel-by-channel and span-by-span power optimization for C+L and S+C+L fixed-grid, fixed-modulation-format wavelength division multiplexing (WDM) links in static networks were reported. However, no effort has as yet been made to optimize the launch power in dynamic EONs over the C+L bands. In this paper, we present an algorithm for jointly optimizing the launch power of the entire channel plan over the C+L bands, so that the cumulative generalized signal-to-noise ratio (GSNR) is maximized, and apply it to a network-level study based on modulation-format and distance-adaptive dynamic routing under two different spectrum assignment (SA) strategies: exact first-fit (EFF) and exact last-fit (ELF). The results show that, unlike in the case of C-band networks, ELF achieves better GSNR performance in C+L-band networks.

## II. System Model and GSNR Estimation Tool

A C+L EON is modeled as a graph with vertices and edges, i.e. *(N,E)*. The vertices are the network nodes equipped with transponders and reconfigurable add-drop multiplexers operating in the C+L bands [1]. An edge between two vertices is the link between those two nodes. If a link is longer than a given average span length, in-line C+L erbium-doped fiber amplifiers (EDFA) equipped with gain-flattening equalizers are considered over that link. In this regard, the end-to-end GSNR of a channel with launch power $P_{ch}$ propagating along a given lightpath (LP) can be estimated as follows [10],

$$GSNR^{i,r-1} \approx \sum_{l=1}^{L^r} \sum_{s=1}^{S^{r,l}} \left( \frac{P_{ch}^{i,l,s} - (P_{ch}^{i,l,s})^3 \eta_{ch}^{i,l,s}}{P_{ASE}^{i,l,s} + (P_{ch}^{i,l,s})^3 \eta_{ch}^{i,l,s}} \right)^{-1} + \left( \kappa^i P_{ch}^{i,l,1} \right)^{-1} \quad (1)$$

where $i$, $r$, $l$ and $s$ are the indices of the $i^{th}$ channel, $r^{th}$ LP, $l^{th}$ link, and $s^{th}$ span, respectively, whereas $\kappa^i$, $\eta_{ch}^{i,l,s}$, $P_{ch}^{i,l,s}$ and $P_{ASE}^{i,l,s}$ are the transceiver noise coefficient (=1/$SNR_{TRX}$) [8], the non-linear-interference (NLI) coefficient, the launch power for span $s$, i.e. $P_{ch}^{i,l,s} = P_{ch}^{i,l,s}(z^{l,s}=0)$, $z^{l,s}$ being the propagation distance; and the noise power of the EDFAs in each span, respectively. The ASE power is derived as $P_{ASE}^{i,l,s} \approx hf^i N_F^{l,s} G^{l,s}$, where $N_F^{l,s}$ and $G^{l,s}$ are the noise figure, and the amplifier gain of span $s$ on link $l$, respectively. $f^i$ is the center frequency offset of channel $i$. We assume in-line amplifiers compensating for the total loss of each span, with non-identical attenuation values equal to $P_{ch}^{i,l,s}(L^{l,s})/P_{ch}^{i,l,s}$, where $L^{l,s}$ is the span length, and $P_{ch}^{i,l,s}(L^{l,s})$ (i.e., received power at the end of span $s$) and $\eta_{ch}^{i,l,s}$ can be obtained from equations (2) and (3) in [9].

## III. Power Optimization Strategy for Dynamic C+L EONs

For C-band fiber transmission, power attenuation can be assumed to be frequency independent so that the power profile can be modeled as an exponential function of $-\alpha z$, where $z$ indicates the transmission distance. This is not the case, as seen from equations (2) and (3) in [9], for C+L band transmission, where the power profile also depends on the channel frequency due to inter-channel SRS (ISRS). Therefore, the maximum reach distance and the optimum injected power are frequency dependent. In Fig.1a-d these facts have been illustrated, assuming a uniform power profile, for links with 70-km spans of ITU-T G.652.D fiber with the parameters given in [1], and EDFAs with $N_F^{l,s}$ = 4.5 and 6 dB in the C and L bands [1], respectively. The Raman gain slope has been considered $28 \times 10^{-3}$ [1/W/Km/THz] [9]. The other required parameters have been clarified in section IV. Due to ISRS, the power from the higher frequency components of the aggregated WDM

978-3-903176-54-6 © 2023 IFIP

Farhad Arpanaei acknowledges support from the CONEX Plus programme funded by Universidad Carlos III de Madrid and the European Union's Horizon 2020 research and innovation programme under the Marie Sklodowska-Curie grant agreement No. 801538. The authors would like to acknowledge the support of the EU-funded B5G-OPEN project (grant no. 101016663), and the Spanish project ACHILLES (PID2019-104207RB-I00).

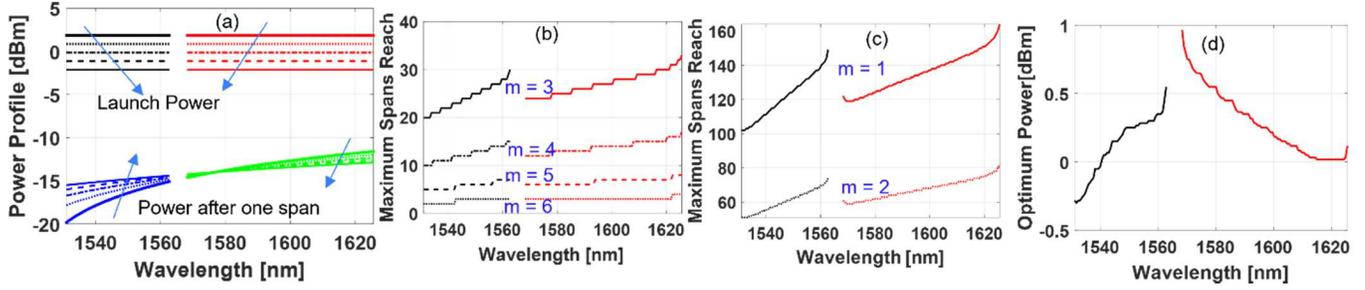

Fig. 1. Shown are (a) the power profile for uniform launch powers 2, 1, 0, -1, and -2 dBm after one span; (b) and (c) maximum reach distance (MRD), expressed as number of spans, for MFC $m$=1-6; and (d) average channel-by-channel optimum powers for MRDs across MFCs [as shown in (b) and (c)], as a function of the channel wavelength, for a span length equal to 70 km. $N_F^{l,s}$= 4.5 and 6 dB in C and L, respectively.

signal is transferred to the lower frequency components, giving rise to a seesaw-shaped power profile as shown in Fig.1a, which we refer to as "seesaw effect" in this paper. The center of gravity of the seesaw shape is the center of the C+L bands.

Therefore, the maximum reach distance (MRD) and the optimum power for each channel and MFC varies in terms of the channel frequency, as shown in Fig.1b-d. Since L-band channels increase their power at the expense of C-band channels as the signal propagates, the MRD for all MFCs in the L band shows a positive offset that decreases as the channel wavelength increases (higher tilt for C-band than L-band). As frequency decreases, maximum reach distance grows stepwise.

In a dynamic C+L EON planning scenario, the strategy described above to optimize the channel power on a channel-by-channel basis is not efficient from an operational and computation-time standpoint, and does not prevent the occasional outage of channels with higher MFC. To find the optimum launch power of all channels across the C+L bands, we propose instead the non-convex optimization algorithm in equations (2-a)-(2-c) below. The objective function of the optimization algorithm is maximizing the cumulative GSNR of all channels (2-a), provided that two constraints (**C1-C2**) are met. **C1** states that the GSNR of a channel $i$ with MFC $m$ must exceed the SNR threshold for its MFC ($SNR_{th}^m$) (2-b). For simplicity, we consider uniform launch power at the beginning of each span ($P_{ch}^{i,l,s} = P_{ch}$), but the algorithm can be extended to tilt-aware power optimization, as proposed in [3,6].

$$\max_{P_{ch}} \sum_{m=1}^{M} \sum_{i=1}^{N_{ch}} GSNR^{i,m}|_{N_{span}^{l,m}}, \quad (2\text{-a})$$

**C1:** $GSNR^{i,m} \geq SNR_{th}^m, \; \forall \; m \in \{1,2,3,4,5,M=6\}, \; i \in [1, N_{ch}]$ (2-b)
**C2:** $P_{ch}^{min} \leq P_{ch} \leq P_{ch}^{max}.$ (2-c)

Additionally, the constraint **C2** reduces the searching space of the particle swarm optimization solution [8] that is considered to solve the proposed optimization problem (2-c). $P_{ch}^{min}$, $P_{ch}^{max}$, and $N_{span}^{l,m}$ are the boundary values of the channel-by-channel optimization strategy illustrated in Fig.1b-d, where $N_{span}^{l,m}$ is the maximum number of spans for the shorter wavelength channel (channel 1) in C-band with MFC $m$, as shortest MRD of each MFC occurs in the first C-band channel (worst case).

## IV. SIMULATIONS AND RESULTS

Following the system model described above, we calculate the optimal channel power and carry out a network level analysis over a C+L EON based on an Italian 21-node, 36-link regional optical backbone network [10], with average span length of 70 km (ITU-T G.652.D fiber with parameters in [1]). We assume coherent super-channel (SpCh) transceivers with sub-channels (SbCh) operating at a fixed baud rate of 64 GBaud with $SNR_{TRX}$ = 36 dB [3] and spectral bandwidth equal to six frequency slots (i.e., 6×12.5 = 75 GHz). Similar to [6], the guard band between the C and L bands is set to 500 GHz. Moreover, we consider a 2-dB aging SNR margin, $N_F^{l,s}$ is fixed at 4.5 and 6 dB in C and L band, respectively; and $BER_{threshold}$ = 1×10$^{-3}$ is considered suitable for a 28% forward error correction overhead. With these values, the GSNR thresholds for MFC $m$=1-6 (corresponding to SbCh bit rates of 100, 200, 300, 400, 500, and 600 Gb/s, respectively) are 6.79, 9.81, 13.71, 16.54, 19.58, and 22.54 dB, respectively. If the requested bit rate exceeds the maximum capacity of a SbCh for a given MFC, the request is established over $N$ adjacent SbChs using the same MFC, forming a SpCh with spectral occupancy of $N$ × 75 GHz. To find the maximum reach distance lookup table and the optimum launch power for each MFC, the optimization algorithm expressed in equation (2) is solved.

Assuming fully loaded links, results show that the optimum launch power is -0.15 dBm, and the MRDs are 102, 51, 20, 10, 5, and 2 spans for $m$ = 1 to 6, respectively. Fig. 2 illustrates the effects of power deviation from the optimum launch power for $m$ = 4 (16QAM). As shown in Fig. 2.a and Fig. 2.c, the seesaw effect appears in NLI coefficient and ASE noise power, but in opposite directions in the same band. As expected, by decreasing the launch power from the optimum value, the NLI noise power decreases in both bands (see Fig. 2.b). Because the launch power decreases, the GSNR declines as well (see Fig. 2.d).

Likewise, when the launch power is set to a value greater than the optimum power, the GSNR decreases because of the increase of the ASE noise in the C band and of the NLI coefficient in the L band. Indeed, the ASE noise in the C band and the NLI noise in the L band have a prominent effect on GSNR. The behavior shown Fig. 2 for PM-16QAM can be extended to all MFCs. The GSNR of all channels with the optimum launch power indicated above is depicted in Fig. 3.a. The GSNR of all channels satisfies the SNR threshold for each MFC. Having calculated the optimum power and MRDs, we do a network-level study in which traffic requests are generated randomly, with bit rates in the range [100,600] Gb/s, with a 100 Gb/s step.

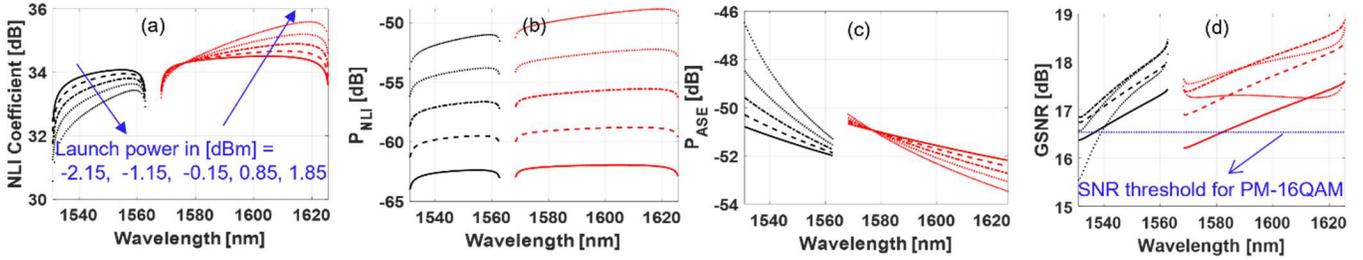

Fig. 2. (a) Non-linear (NLI) coefficient, (b) NLI noise power, (c) ASE noise power, and (d) GSNR all in dB over 10×70 km with launch power -2.15, -1.15, -0.15, 0.85, 1.85 [dBm] with PM-16QAM.

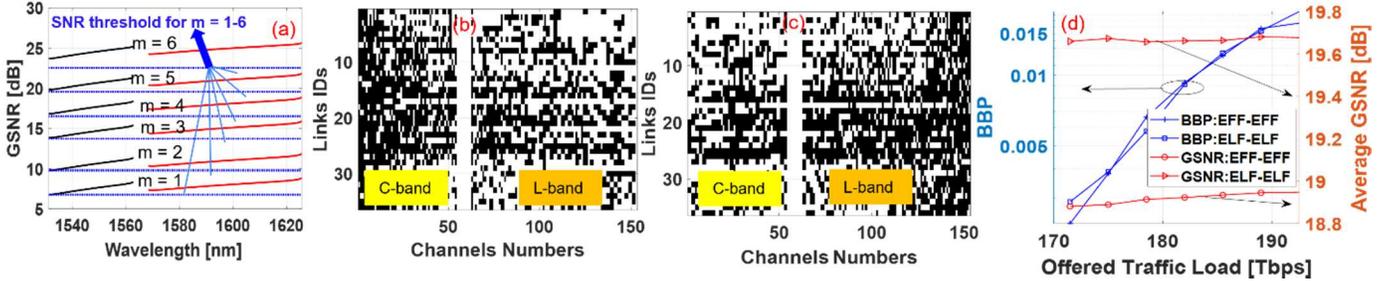

Fig. 3. (a) GSNR in dB for PM-BPSK, PM-QPSK, PM-8,16, 32, and 64QAM labeled by m=1-6, respectively, over 102, 51, 20, 10, 5, and 2 × 70 km with -0.15 dB launch power. (b) and (c) the snapshot spectrum of exact first-fit (EFF) and exact last-fit (ELF) in C and L band (black: busy channel, white: idle channel), (d) bandwidth blocking probability (BBP) and lightpaths' average arrival GSNR.

The requests arrive according to a Poisson distribution with an average arrival rate of AT requests per time unit. We do not impose any restrictions on the BVTs at the nodes. The holding time of each request is generated according to a negative exponential distribution with an average of 1/HT. The offered traffic load (OTL) is AT/HT normalized traffic units. The distribution of requests between source-destination nodes is uniform, i.e., each node is selected as a source or destination node with the same probability.

For each examined value of the OTL, one set of requests is simulated with five repetitions in each OTL. We use $k = 3$ disjoint shortest paths to generate candidate LPs, and consider ASE-noise loading for unused channels to be in accordance with the above assumptions. Two SAs are studied based on the GSNR behavior: (1) EFF in the C band and L band (EFF-EFF), and (2) ELF in the C band and L band (ELF-ELF). Snapshots of the spectrum for the last OTL are shown in Fig. 3.b and Fig. 3.c, respectively. As seen in Fig. 3.d, the bandwidth blocking probability (BBP) is the same for both SAs, but the LP's average GSNR at the arrival time in ELF-ELF is from 0.9 dB to 0.7 dB higher than in EFF-EFF for lower OTL to higher OTL.

## V. CONCLUSION

In this paper, we showed that channel-by-channel power optimization is not adequate for dynamic C+L EONs, not only on account of computation time and operation complexity, but also because it does not guarantee that the SNR threshold requirement is met for all channels and MCFs. To solve this problem, we proposed a launch power optimization algorithm that maximizes the cumulative GSNR of all channels across the C+L bands while ensuring the GSNR of each channel exceeds the SNR threshold. The application of this algorithm to a dynamic EON planning study revealed that the exact last fit (ELF) spectrum assignment (SA) outperforms the exact first fit (EFF) SA in terms of average GSNR at arrival time (i.e. the SNR design margin can be reduced). This result is unlike in C-band SA strategies, where both ELF and EFF perform equally well, and can be explained by the positive tilt that exhibits the GSNR for decreasing frequencies in C+L fiber transmission.